\documentclass[preprints,article,accept,moreauthors,pdftex]{Definitions/mdpi} 

\usepackage{graphicx}
\usepackage{subcaption}
\usepackage{comment}
\usepackage{enumitem}
\firstpage{1} 
\pubvolume{xx}
\issuenum{1}
\articlenumber{5}
\pubyear{2019}
\copyrightyear{2019}
\history{Received: date; Accepted: date; Published: date}

\Title{Including arbitrary geometric correlations into one-dimensional time-dependent Schr\"{o}dinger equations}

\Author{Devashish Pandey $^{1,*}$, Xavier Oriols $^{1,*}$ and Guillermo Albareda $^{2,3,*}$}

\AuthorNames{Devashish Pandey, Xavier Oriols and Guillermo Albareda}

\address{%
$^{1}$ Departament d'Enginyeria Electr\`onica. Universitat Aut\`onoma de Barcelona. Edifici Q. 08193 Bellaterra. \\
$^{2}$ Max Planck Institute for the Structure and Dynamics of Matter and Center for Free-Electron Laser Science, Luruper Chaussee 149, 22761 Hamburg, Germany. \\
$^3$ Institut de Qu\'{i}mica Te\`orica i Computacional (IQTCUB), Universitat de Barcelona, Mart\'{i} i Franqu\`es 1, 08028 Barcelona, Spain.}

\corres{Correspondence: devashish.pandey@uab.cat (D.P), xavier.oriols@uab.cat (X.O), guillermo.albareda@mpsd.mpg.de (G.A)}

\abstract{The so-called Born-Huang ansatz is a fundamental tool in the context of ab-initio molecular dynamics, viz., it allows to effectively separate fast and slow degrees of freedom and thus treating electrons and nuclei at different mathematical footings.
Here we consider the use of a Born-Huang-like expansion of the three-dimensional time-dependent Schr\"odinger equation to separate transport and confinement degrees of freedom in electron transport problems that involve geometrical constrictions. The resulting scheme consists of an  eigenstate problem for the confinement degrees of freedom (in the transverse direction) whose solution constitutes the input for the propagation of a set of coupled one-dimensional equations of motion for the transport degree of freedom (in the longitudinal direction). This technique achieves quantitative accuracy using an order less computational resources than the full dimensional simulation for a prototypical two-dimensional constriction. 
}


\begin{document}

\section{Introduction}

Nanoscale constrictions (sometimes referred to as point contacts or nanojunctions) are unique objects for the generation and investigation of ballistic electron transport in solids. Studies of such systems have been inspired by the pioneering investigations by Sharvin in the mid-1960s \cite{sharvin1965possible}. 
Today, advances in the fabrication techniques like direct growth of branched nanostructures \cite{jin2006direct}, electron beam irradiation \cite{terrones2002molecular}, thermal and electrical welding \cite{peng2009bottom} or atomic force microscope \cite{shen2009nondestructively} have allowed to control the size and composition of nanojunctions for creating devices with desired functionalities. 
In this respect, a number of nanodevices based on nonjunctions like the single electron transistors \cite{takahashi1995fabrication,maeda2012logic}, field effect transistors \cite{tans1998room,zhang2008assessment} and heterostructure nanowires \cite{nah2010hbox,hu2007ge} have been recently reported which promise great performance in terms of miniaturization and power consumption. 

In the design of these nanostructures, simulation tools constitute a valuable alternative to the expensive and time-consuming test-and-error experimental procedure. 
Today, a number of quantum electron transport simulators are available to the scientific community~\cite{nemo,nextnano,tibercad,nanocad,transiesta}. The amount of information that these simulators can provide, however, is mainly restricted to the stationary regime and therefore their predicting capabilities are still far from those of the traditional Monte Carlo solution of the semi-classical Boltzmann transport equation~\cite{jacoboni1983monte}. This limitation poses a serious problem in the near future as electron devices are foreseen to operate at the Terahertz (THz) regime. At these frequencies, the discrete nature of electrons in the active region is expected to generate unavoidable fluctuations of the current that could interfere with the correct operation of such devices both for analog and digital applications~\cite{Albareda_2009}. 

A formally correct approach to electron transport beyond the quasi-stationary regime lies on the modeling of the active region of electron devices as an open quantum system~\cite{breuer2002theory,smirne2010initial}.
As such, one can then borrow any state-of-the-art mathematical tool developed to study open quantum systems~\cite{de2017dynamics,PhysRevA.78.022112}. A preferred technique has been the stochastic Schr\"{o}dinger equation (SSE) approach~\cite{gisin1989stochastic,pearle1989combining, carmichael2009open,van1992stochastic,de2011non,goetsch1994linear,gatarek1991continuous,gambetta2002non}. Instead of directly solving equations of motion for the reduced density matrix, the SSE approach exploits the state vector nature of the so-called conditional states to alleviate some computational burden~\cite{rivas2014quantum}.

An example of the practical utility of the SSE, a Monte Carlo simulation scheme to describe quantum electron transport in open systems that is valid both for Markovian or non-Markovian regimes and that guarantees a dynamical map that preserves complete positivity has been recently proposed~\cite{e21121148}.
The resulting algorithm for quantum transport simulations reformulates the traditional "curse of dimensionality'' that plagues all state-of-the-art techniques for solving the time-dependent Schr\"odinger equation (TDSE). Specifically, the algorithm consists on the solution of an ensemble of single-particle SSEs that are coupled, one to each other, through effective Coulombic potentials~\cite{PhysRevLett.98.066803,albareda2009many}. Furthermore, the simulation technique 
accounts for dissipation~\cite{colomes2017quantum} and
guarantees charge and current conservation through the use of self-consistent time-dependent boundary conditions~\cite{albareda2010time,albareda2013self,BITLLES8} that partially incorporate the exchange interaction~\cite{BITLLES7,albareda2016electron}. Solving a large number of three-dimensional (3D) single-particle TDSEs, however, may still be a very time-consuming task. Therefore, the above technique would greatly benefit from the possibility of further reducing the dimensionality of the numerical problem.

It is the purpose of this work to derive and discuss a method that allows to solve the 3D TDSE in terms of an ensemble of one-dimensional (1D) TDSEs. The technique is inspired on the so-called Born-Huang ansatz~\cite{born1954dynamical}, which is a fundamental tool in the context of ab-initio molecular dynamics that allows to separate fast and slow degrees of freedom in an effective way~\cite{PhysRevA.94.062511}. Here we consider an analogous ansatz to separate transport and confinement directions. As it will be shown, the resulting technique allows us to include arbitrary geometric correlations into a coupled set of 1D TDSEs. Therefore, while we have motivated the development of this method in the context of the simulation of (non-Markovian) quantum transport in open system, the method presented here could be of great utility in many research fields where the reduction of the dimensionality in quantum systems with geometrical correlations may be advantageous. 

The manuscript is structured as follows. In Section \ref{non-adiabatic_eq} we introduce a Born-Huang-like ansatz that allows to expand the 3D single-particle TDSE in terms of an infinite set of (transverse) eigenstates weighted by (longitudinal) complex coefficients. The equations of motion for the coefficients are found to be coupled and obey a linear (non-unitary) partial differential equation.
In Section~\ref{ApplicationI} we apply the method to a prototypical 2D constriction. Section \ref{application} is devoted to find analytical expressions for the effective potentials that appear in the equation of motion of the coefficients. 
A discussion on the geometrical dependence of these effective potentials is provided. In Section \ref{numerical_example} we illustrate the performance of the method to describe the dynamics of an electron across the 2D nanojunction. In Section \ref{GD_C}
we provide a thorough discussion on the advantages and potential drawbacks of the method. 
We conclude in Section \ref{conclusions}.

\section{Single-electron time-dependent Schr\"odinger equation in a Born-Huang-like basis expansion}
\label{non-adiabatic_eq}

As we have explained in the introduction, it is our goal to reduce the computational burden associated to the solution of an ensemble of effective single-electron 3D SSE~\cite{e21121148}. 
Therefore, we consider our starting point to be the 3D TDSE of a single (spin-less) electron in the position basis, i.e.:
\begin{equation}\label{com}
	 i \frac{\partial}{\partial t}\Psi(x,y,z,t)= H(x,y,z)\Psi(x,y,z,t),
\end{equation}
where we have used atomic units, and $x$, $y$ and $z$ represent the three spatial coordinates. 
In Equation~\ref{com},  
$H(x,y,z)$ is the full Hamiltonian of the system:
\begin{eqnarray}
\label{H}
    H(x,y,z) &=& T_x + T_y + T_z + V(x) + W(x,y,z),
\end{eqnarray}
which has been assumed to be time-independent for simplicity.
The time-dependence on the scalar potentials $V(x)$ and $W(x,y,z)$ will be discussed in later sections.
In Equation~\ref{H}, $T_x=-\frac{1}{2}\frac{\partial^2}{\partial x^2}$ and $V(x)$ are, respectively, the kinetic energy and the scalar potential associated to the longitudinal degree of freedom $x$, while $T_y=-\frac{1}{2}\frac{\partial^2}{\partial y^2}$ and $T_z=-\frac{1}{2}\frac{\partial^2}{\partial z^2}$ are the kinetic energies associated to the transversal degrees of freedom $y$ and $z$.
The scalar potential $W(x,y,z)$ includes any other scalar potential that is not purely longitudinal, which is responsible of making the solution of Equation~\ref{com} non-separable. 

It is convenient at this point to rewrite the Hamiltonian in Equation~\ref{H} in terms of longitudinal and transverse components as:
\begin{equation}
H(x,y,z) = T_x + V(x) + H^\perp_x(y,z),
\label{H_tot}
\end{equation}
where $H^\perp_x(y,z)$ is the transverse Hamiltonian defined as:
\begin{equation}
    H^\perp_x(y,z) = T_y + T_z + W(x,y,z).
    \label{transverse_hamiltonian}
\end{equation} 
An eigenvalue equation associated to the transverse Hamiltonian can now be introduced as follows:
\begin{eqnarray}
	H^\perp_x(y,z)\phi^k_x(y,z)=\mathcal{E}^k(x)\phi^k_x(y,z),
\label{incom}
\end{eqnarray}
where $\mathcal{E}^k(x)$ and $\phi^k_x(y,z)$ are the corresponding eigenvalues and eigenstates respectively. The eigenstates $\phi^k_x(y,z)$ form a complete basis in which to expand the Hilbert space spanned by the variables $x$, $y$, and $z$. Therefore, the 3D wavefunction in Equation~\ref{com} can be expressed in terms of transverse eigenstates $\phi^k_x(y,z)$ as:
\begin{eqnarray}\label{wf}
    \Psi(x,y,z,t)=\sum_{k=1}^\infty\chi^k(x,t)\phi^k_x(y,z),
\end{eqnarray}
where $\chi^k(x,t) = \iint dydz \phi^k_x(y,z)\Psi(x,y,z,t)$ are complex longitudinal coefficients associated to the transverse eigenstate $\phi_x^k(y,z)$. Unless otherwise stated all integrals are evaluated from $-\infty$ to $\infty$. 
It is important to note that since the longitudinal variable $x$ appears as a parameter in Equation~\ref{incom}, the transverse eigenstates obey the following partial normalization condition:
\begin{equation}\label{partial_norm}
    \iint dydz \phi^l_x(y,z) \phi^k_x(y,z) = \delta_{lk}, \;\; \forall x.
\end{equation}
In addition, the longitudinal complex coefficients $\chi^k(x,t)$ in Equation~\ref{wf} fulfill, by construction, the condition:
\begin{equation}\label{norm}
    \sum_{k=1}^\infty \int dx |\chi^k(x,t)|^2 = 1.
\end{equation}
The wavefunction expansion in Equation~\ref{wf} can now be introduced into Equation~\ref{com} to obtain an equation of motion for the coefficients $\chi^k(x,t)$ (see Appendix \ref{non_adiabatic_eq_derivation}):
\begin{equation}
	i\frac{\partial}{\partial t}\chi^k(x,t)=\left(T_x+\mathcal{E}^k(x)+V(x)\right)
	\chi^k(x,t) - \sum_{l=1}^\infty\left(S^{kl}(x) +F^{kl}(x)\frac{\partial}{\partial x}\right)\chi^l(x,t),
	\label{nadse}
\end{equation}
where $\mathcal{E}^k(x)$ are \textit{effective potential-energies} (that correspond to the eigenvalues in Equation~\ref{incom}) and $F^{kl}(x)$ and $S^{kl}(x)$ are geometric (first and second order) coupling terms, which read:
\begin{subequations}\label{NACS}
\begin{equation}
    F^{kl}(x)=\iint dy dz \phi^{*l}_x(y,z)\frac{\partial}{\partial x} \phi^k_x(y,z),
\end{equation}
\begin{equation}
    S^{kl}(x)=\frac{1}{2} \iint dy dz \phi^{*l}_x(y,z)\frac{\partial^2}{\partial x^2}\phi^k_x(y,z).
\end{equation}
\end{subequations}
Since the transverse eigenstates $\phi^k_x(y,z)$ are real, the term $F^{kk}$ is zero by construction. The other terms in Equation~\ref{NACS} dictate the transfer of probability presence between different longitudinal coefficients $\chi^k(x,t)$ and, therefore, will be called \textit{geometric non-adiabatic couplings} (GNACs).
Accordingly, one can distinguish between two different dynamics regimes in Equation~\ref{nadse}:  
\begin{enumerate}[label=(\roman*)]
    \item \textbf{Geometric adiabatic regime}: it is the regime where $F^{kl}$ and $S^{kl}$ are both negligible. Thus, the solution of Equation~\ref{nadse} can be greatly simplified because it involves only one transverse eigenstate.
    \item \textbf{Geometric non-adiabatic regime}: it is the regime where either or both $F^{kl}$ and $S^{kl}$ are important. Thus, the solution of Equation~\ref{nadse} involves the coupling between different longitudinal coefficients and hence more than one transverse eigenstate.
\end{enumerate}
Interestingly, the prevalence of the regimes (i) or (ii) can be estimated by rewriting the first order coupling terms $F^{kl}(x)$ as (see Appendix \ref{adiabatic_condition} for an explicit derivation):
\begin{equation}
    F^{kl}(x)=\frac{\int dy \int dz \phi^{*l}_x(y,z)\left(\frac{\partial}{\partial x}W(x,y,z)\right) \phi^k_x(y,z)}{\mathcal{E}^l(x)-\mathcal{E}^k(x)} \;\;\;\; \forall k \neq l.
    \label{adiabatic_con}
\end{equation}
That is, the importance of non-adiabatic transitions between transverse eigenstates depends on the interplay between the transverse potential-energy differences $\mathcal{E}^l(x)-\mathcal{E}^k(x)$ and the magnitude of the classical force field $\propto\frac{\partial}{\partial x}W(x,y,z)$. 
The geometric adiabatic regime (i) is reached either when the classical force field is very small or the energy differences $\mathcal{E}^l(x)-\mathcal{E}^k(x)$ are large enough. In the adiabatic regime only the diagonal terms, $S^{kk}$, are retained,
which induce a global shift of the potential-energies $\mathcal{E}^k(x)$ felt by the longitudinal coefficients $\chi^k(x,t)$.
In this approximation, the longitudinal degree of freedom moves in the potential-energy provided by a single transverse state, $\mathcal{E}^k(x)$. 
This regime is analogous to the so-called \textit{Born-Oppenheimer approximation} in the context of molecular dynamics~\cite{Leticia_book}, where the term $S^{kk}$ is often called Born-Oppenheimer diagonal correction~\cite{handy1986diagonal}.
As it will be shown in our numerical example, the evolution of the system can be governed either by the geometric adiabatic or nonadiabatic regime depending on the particular spatial region where the dynamics is occurring.

Let us notice at this point that the time-dependence of the Hamiltonian in Equation~\ref{H} may come either due to a purely longitudinal time-dependent scalar potential $V(x,t)$ or through the time-dependence of the non-separable potential $W(x,y,z,t)$. If the time dependence is added only through $V(x,t)$, then nothing changes in the above development. Contrarily, if a time-dependence is included in $W(x,y,z,t)$, then the eigenstate problem in Equation~\ref{incom} changes with time and so do the effective potential-energies $\mathcal{E}^k(x,t)$ and the first and second order GNACs $F^{kl}(x,t)$ and $S^{kl}(x,t)$. As it will be shown later, in this circumstance, Equation~\ref{incom} should be solved self-consistently with Equation~\ref{nadse}.

Before we move to a practical example implementing the above formulation of the 3D TDSE, let us emphasize that it is the main goal of the set of coupled equations in Equation~\ref{nadse} to allow the evaluation of relevant observables in terms of 1D wavefunctions only. In this respect, let us take, for example, the case of the reduced probability density $\rho(x,t) = \iint dy dz \Psi^*(x,y,z,t)\Psi(x,y,z,t)$.
Using the basis expansion in Equation~\ref{wf}, $\rho(x,t)$ can be written as:
\begin{equation}
\rho(x,t)=\sum_{k,l}^{\infty}\chi^{*l}(x,t)\chi^k(x,t)\iint dy dz \phi^{*l}_x(y,z)\phi^k_x(y,z),
\label{reduced_density2}
\end{equation}    
and using the condition in Equation~\ref{partial_norm} the above expression reduces to:
\begin{equation}
\rho(x,t)= \sum_{k=1}^{\infty}|\chi^{k}(x,t)|^2.
\label{rho2d}
\end{equation}
Therefore, according to Equation~\ref{rho2d}, the reduced (longitudinal) density is simply the sum of the absolute squared value of the longitudinal coefficients $\chi^k(x,t)$, which is in accordance with Equation~\ref{norm}, i.e., $\int dx \rho(x,t) = \sum_{k=1}^{\infty} \int dx |\chi^{k}(x,t)|^2 = 1$.
Similarly, other relevant observables, such as the energy can easily be derived using the expansion in Equation~\ref{wf} (see Appendix~\ref{mean_energy}).

\section{Application of the method to a prototypical constriction}
\label{ApplicationI}
The above formulation can be cast in the form of a numerical scheme to solve the 3D TDSE, which we will call, hereafter, geometrically correlated 1D TDSE (or in brief GC-TDSE). 
The scheme can be divided into two different parts corresponding to their distinct mathematical nature. The first part involves the solution of the eigenvalue problem in Equation~\ref{incom} which allows to evaluate the transverse eigenstates $\phi^k_x(y,z)$ and eigenvalues $\mathcal{E}^k(x)$ as well as geometric non-adiabatic couplings $F^{kl}(x)$ and $S^{kl}(x)$ in Equation~\ref{NACS}. These quantities are required in the second part of the algorithm to solve the equation of motion of the longitudinal coefficients $\chi^k(x,t)$ in Equation~\ref{nadse}, which ultimately allow us to evaluate the observables of interest.

In what follows we discuss these two aspects of algorithm for a prototypical 2D geometric constriction whose geometry does not change in time.
We consider one degree of freedom in the transport direction and one degree of freedom in the transverse (or confinement) direction. The generalization to a 3D system, i.e., with two transverse degrees of freedom, is straightforward and does not add any physical insight to the 2D case. As it will be shown, the transverse eigenstates and eigenvalues as well as the geometric non-adiabatic couplings are, for a time-independent constriction, functions that can be computed once and for all. That is, the effective potential-energies $\mathcal{E}^k(x)$ and the non-adiabatic couplings $F^{kl}$ and $S^{kl}$ are computed only at the very beginning of the GC-TDSE propagation scheme. For more general time-dependent constrictions, possibly with no analytical form of $W(x,y,z)$ and hence of $\phi_k(y)$ and $\mathcal{E}^k(x)$, the only change in the algorithm is that Equation~\ref{incom} has to be solved, self-consistently, together with Equation~\ref{nadse} at each time step.

\subsection{Evaluation of transverse eigen-states (and values) and geometric non-adiabatic couplings}	
\label{application}

Let us consider the case of a 2D nanojunction represented by the scalar potential:
\begin{equation}\label{potential}
    V(x,y) =  
\begin{cases}
    0,& \text{if }\; L_2(x)<y< L_1(x)  \\
    \infty,                  & \text{otherwise}
\end{cases}
\end{equation}
where $L_1(x)$ and $L_2(x)$ define the shape of the constriction. 
Given the 2D Hamiltonian,
\begin{equation}\label{appl_ham}
    H(x,y) = T_x + V(x) + H^\perp_x(y)
\end{equation}
where $H^\perp_x(y) = T_y + W(x,y)$. The wavefunction for a 2D constriction in terms of Born-Huang expansion can be written as, $\Psi(x,y,t)=\sum_k\chi^k(x,t)\phi^k_x(y)$,
and the transverse states $\phi^k_x(y)$ are solutions of a free particle in a 1D box whose length depends upon the longitudinal variable $x$, i.e.:
\begin{equation}\label{eig_states}
\phi_x^k(y)=
\begin{cases}
	\sqrt{\frac{2}{L(x)}}\mbox{sin}\left(\frac{k\pi (y-L_2(x))}{L(x)}\right),& \text{if }\; L_2(x)<y< L_1(x) \\
	0, & \text{otherwise}.
\end{cases}
\end{equation}
The associated eigenvalues are given by:
\begin{equation}\label{eig_energies}
	\mathcal{E}^k(x) = \frac{k^2\pi^2}{2L^2(x)},
\end{equation}
where we have defined $L(x)=L_1(x)-L_2(x)$.
These energies, parametrically dependent on the longitudinal variable $x$, define the effective potential-energies where the coefficients $\chi^k(x,t)$ evolve on. 
\begin{figure}
\centering
\includegraphics[width=0.75\textwidth]{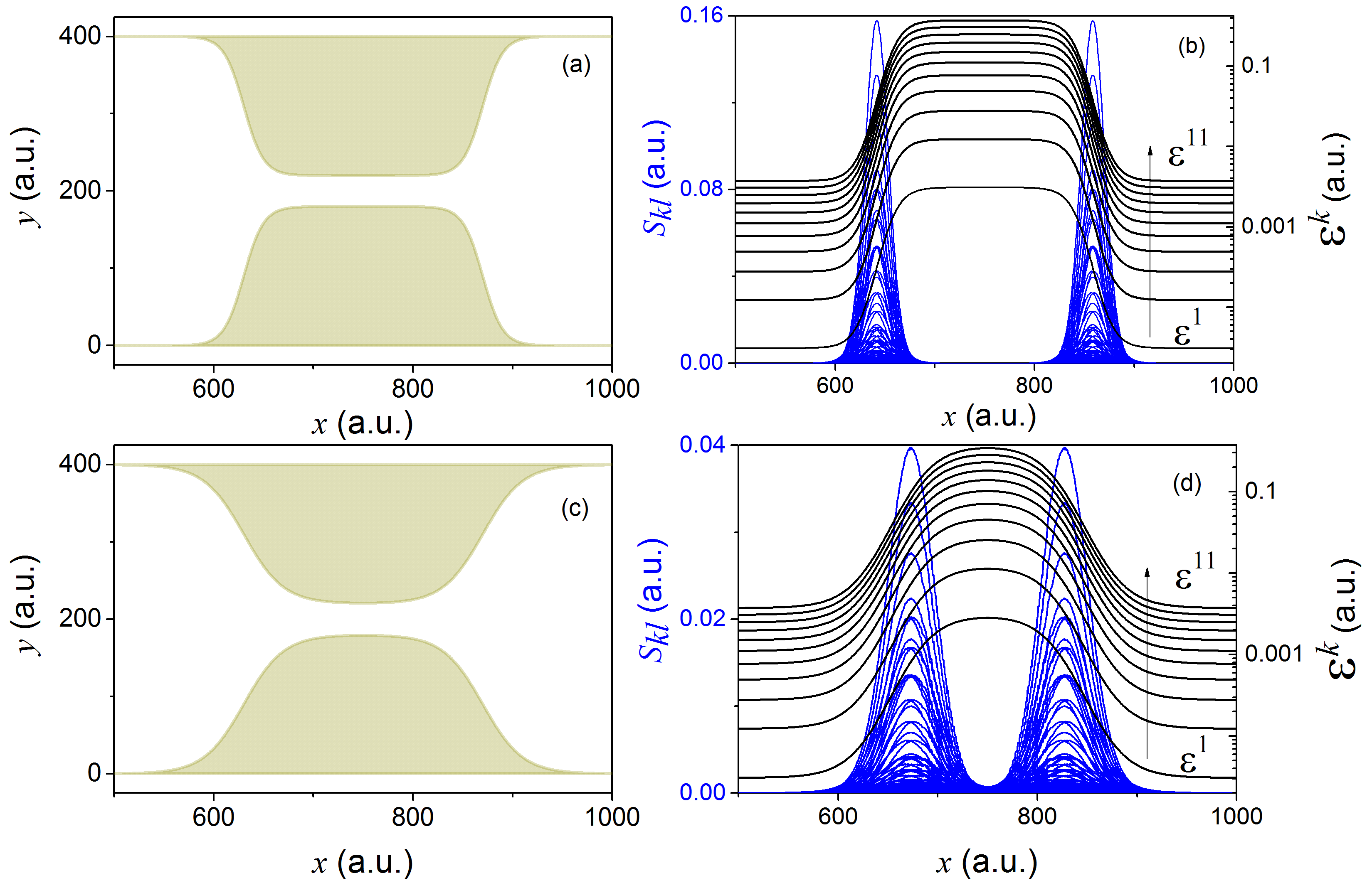}
\caption{Two different nanojunctions, viz., (a) and (c), defined by Equation~\ref{lx1_lx2} in Appendix~\ref{Nanoconstriction_function} and using $\mathcal{A}=180$, $\mathcal{B}=220$, $a_1=630$,  $a_2=870$, with $\gamma=10$ in (a) and $\gamma=20$ in (c). Panels (b) and (d) show the associated second order (non-adiabatic) couplings $S^{kl}$ (solid blue lines) and the associated potential-energies $\mathcal{E}^k(x)$ (solid black lines) for the geometries in (a) and (c) respectively. Note that, due to the symmetry of the states defined in Equation~\ref{eig_states}, the coupling between odd and even states is zero, i.e., $F^{kl}(x) = S^{kl}(x) = 0, \;\forall k+l=\text{odd}$.}
\label{ES}
\end{figure}

To evaluate the first and second order coupling terms $F^{kl}(x)$ and $S^{kl}(x)$, we need to rely on a particular form of the constriction. Depending on the specific form of $L_1(x)$ and $L_2(x)$, different geometrically bounded constrictions can be conceived (see for example the panels (a) and (c) of Figure~\ref{ES}). 
Given the states in Equation~\ref{eig_states} and a particular shape of the constriction (defined in Equation~\ref{lx1_lx2} of Appendix~\ref{Nanoconstriction_function}), it is then easy to evaluate the non-adiabatic coupling terms $F^{kl}$ and $S^{kl}$ (see panels (b) and (d) of Figure~\ref{ES}).

The two different constrictions in Figure~\ref{ES} serve well to gain some insight into the general form and dependence of the effective potential-energies $\mathcal{E}^k(x)$ as well as of the GNACs in Equation~\ref{NACS}.
Geometries changing more abruptly lead to sharper effective potential-energies $\mathcal{E}^k(x)$ and more peaked non-adiabatic coupling terms $F^{kl}(x)$ and $S^{kl}(x)$. 
Sharper constrictions are thus expected to cause larger non-adiabatic transitions and hence to involve a larger number of transverse eigenstates requiring a larger number of longitudinal coefficients in order to reconstruct the reduced (longitudinal) density in Equation \ref{reduced_density2}. On the contrary, smoother constrictions should yield softer non-adiabatic transitions and hence involve a smaller number of transverse eigenstates and equivalently a smaller number of longitudinal coefficients.

\subsection{Time-dependent propagation of the longitudinal coefficients}
\label{numerical_example}
\begin{figure}
\centering
\includegraphics[width=0.6\textwidth]{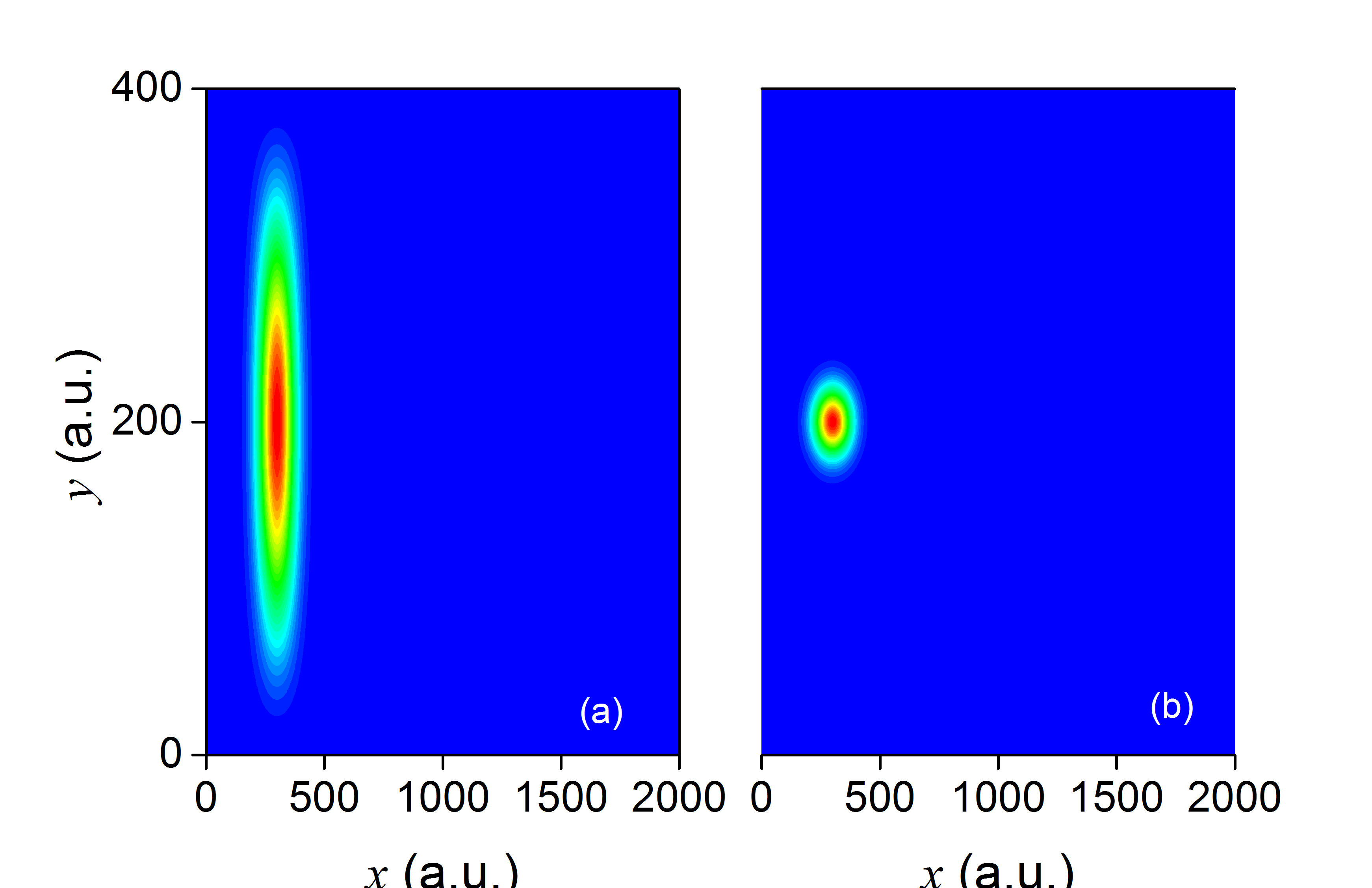}
\caption{Panels (a) and (b) represent the probability density $|\Psi(x,y,0)|^2$, associated to the wavefunctions in Equations~\ref{first_state} and \ref{second_state} respectively. Red regions in the plots correspond to higher probability densities while blue regions correspond to lower probabilities.}
\label{2D_wavefunction}
\end{figure}

Given the effective potential-energies $\mathcal{E}^k(x)$ and the non-adiabatic couplings $F^{kl}(x)$ and $S^{kl}(x)$, one can then easily find a solution for the longitudinal coefficients in Equation~\ref{nadse}. Here, we consider the dynamics of an electron that impinges upon a constriction defined by Equations~\ref{potential} and \ref{lx1_lx2} in Appendix~\ref{Nanoconstriction_function} using the particular set of parameters, $\{\mathcal{A}, \mathcal{B}, a_1, a_2, \gamma\} =\{180,  220, 630, 870, 20\}$, which corresponds to panel (c) of Figure~\ref{ES}.
For this particular set of parameters, the symmetry of the states defined in Equation~\ref{eig_states} forbids transitions between odd and even states, i.e., 
\begin{equation}\label{even_only}
    F^{kl}(x) = S^{kl}(x) = 0\;\;\;\;\forall k+l=\text{odd}.
\end{equation}
We will consider two different initial states. On one hand, the initial wavefunction $\Psi(x,y,0)$ will be described by:
\begin{equation}\label{first_state}
    \Psi(x,y,0) = \phi^{1}_{x}(y)\psi(x),
\end{equation}
where $\phi^{1}_{x}(y)$ is the transverse ground state defined in Equation~\ref{eig_states}, and $\psi(x) = \mathcal{N}\exp\Big(ik_0(x-x_0)\Big)\exp\Big(\frac{-(x-x_0)^2}{2\sigma_x^2}\Big)$ is a minimum uncertainty (Gaussian) wavepacket with initial momentum and dispersion $k_0 = 0.086$ a.u. and $\sigma_x = 80$ a.u. respectively, and centered at $x_0=300$ a.u. (while $\mathcal{N}$ is a normalization constant).
On the other hand, we will consider the initial state to be defined by:
\begin{equation}\label{second_state}
    \Psi(x,y,0) = \xi(y)\psi(x),
\end{equation}
where now both $\xi(y)$ and $\psi(x)$ (defined above) are Gaussian wavepackets. In particular, $\xi(y) =\mathcal{M}\exp\Big(\frac{-(y-y_0)^2}{2\sigma^2}\Big)$ with $y_0 =200$ a.u., $\sigma_y = 20$ a.u. (and $\mathcal{M}$ a normalization constant). The probability densities $|\Psi(x,y,0)|^2$ associated to the above two initial states can be seen, respectively, in panels (a) and (b) of Fig.~\ref{2D_wavefunction}. 
Given the initial states in Equation~\ref{first_state} and Equation~\ref{second_state}, we can then deduce the corresponding longitudinal coefficients as follows:
\begin{equation}
    \chi^k(x,0) = \int dy \phi^k_x(y)\Psi(x,y,0).
\end{equation} 
While the initial state in Equation~\ref{first_state} corresponds to $\chi^k(x,0) = \delta_{k1}\psi(x)$, the state in Equation~\ref{second_state} involves a larger number of transverse eigenstates. Note that this second initial condition may be more realistic in practical situations, as large enough reservoirs may imply a quasi-continuum of transverse states according to Equation~\ref{eig_energies}.    

\begin{figure}
\centering
\includegraphics[width=1\textwidth]{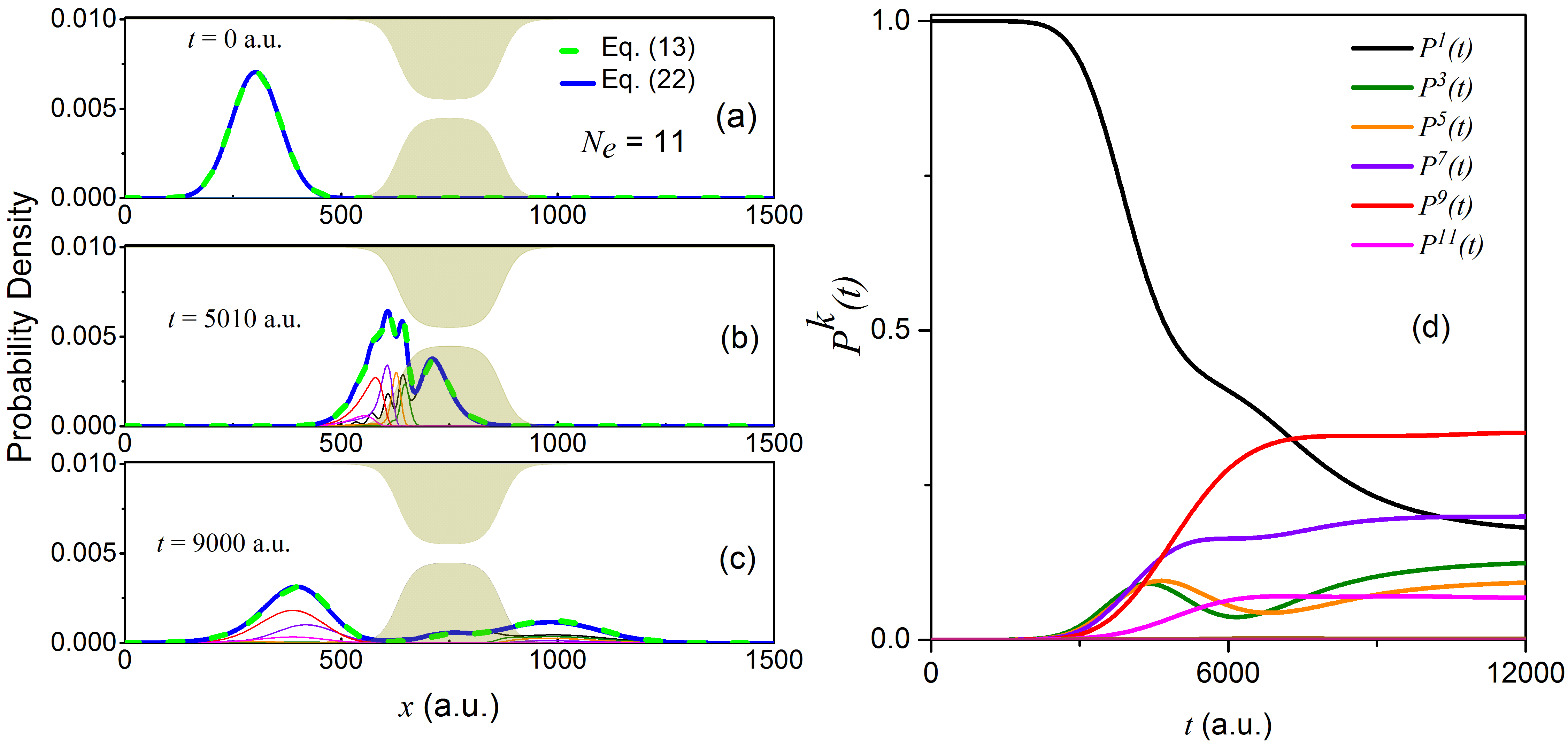}
\caption{Time-evolution of the initial wavefunction in Equation~\ref{first_state}. The reduced density in Equation~\ref{rho2d} (dashed green line) as well as the reduced density in Equation~\ref{rho2d_Ne} for $N_e = 11$ (solid dark blue) are shown at times $t=300$ a.u, $t=5010$ a.u. and $t=9000$ a.u in panels (a), (b) and (c) respectively. The rest of lines correspond to the absolute squared value of the longitudinal coefficients $|\chi^k(x,t)|$.
The evolution of the adiabatic populations in Equation~\ref{populations} can be found in panel (d), using the same color code as in panels (a), (b) and (c).}
\label{evol_ground}
\end{figure}

\begin{figure}
\centering
\includegraphics[width=1\textwidth]{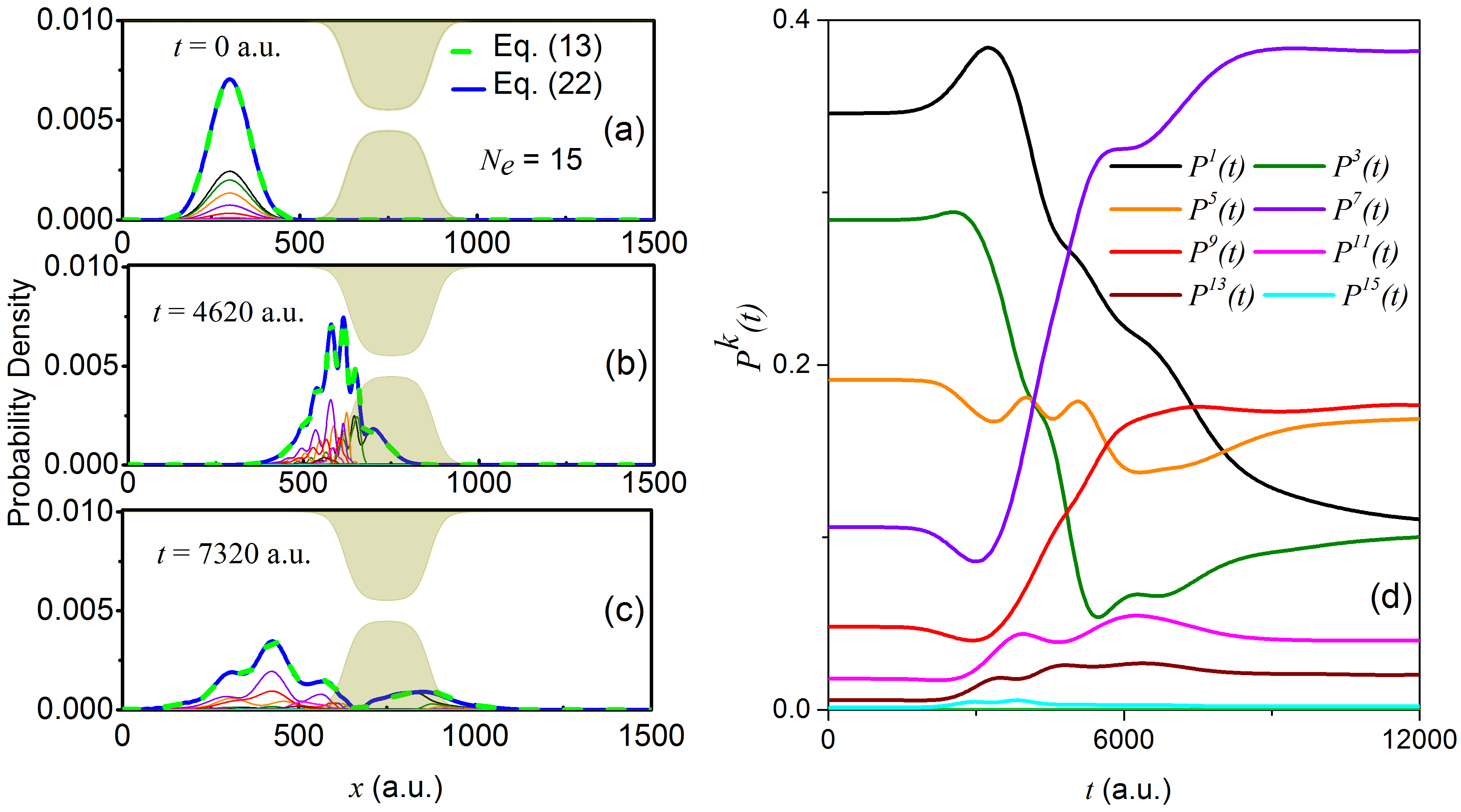}
\caption{Time-evolution of the initial wavefunction in Equation~\ref{second_state}. The reduced density in Equation~\ref{rho2d} (dashed green line) as well as the reduced density in Equation~\ref{rho2d_Ne} for $N_e = 15$ (solid dark blue line) are shown at times $t=0$ a.u, $t=4620$ a.u. and $t=7320$ a.u in panels (a), (b) and (c) respectively. The rest of lines correspond to the absolute squared value of the longitudinal coefficients $|\chi^k(x,t)|$.
The evolution of the adiabatic populations in Equation~\ref{populations} can be found in panel (d), using the same color code as in panels (a), (b) and (c).}
\label{evol_gauss}
\end{figure}

Starting either from Equation~\ref{first_state} or \ref{second_state}, we then propagate the resulting longitudinal coefficients at the initial time according to Equation~\ref{nadse}. Specifically, we used a fourth order Runge-Kutta method with a time-step size of $\Delta t = 0.1$ a.u. and a spatial grid of $1500$ a.u. points with a grid spacing $\Delta x = 1$ a.u. 
In the left panels of Figs.~\ref{evol_ground} and \ref{evol_gauss} we show the time-dependent reduced density of Equation~\ref{rho2d} evaluated from the full 2D wavefunction (dashed green line), as well as the reduced density $\rho(x,t)$ evaluated using the GC-TDSE scheme for a finite number of transversal states $N_e$, i.e:
\begin{equation}
    \rho(x,t) = \sum_{k=1}^{N_e} |\chi^k(x,t)|^2.
    \label{rho2d_Ne}
\end{equation}
In addition, we also show the absolute squared value of the longitudinal coefficients, i.e., $|\chi^k(x,t)|^2$, evaluated using the GC-TDSE. 
Alternatively, in the right panels of Figs.~\ref{evol_ground} and \ref{evol_gauss} we plot the population of each transverse state,
\begin{equation}\label{populations}
    P^k(t) = \int dx|\chi^k(x,t)|^2,    
\end{equation}
as a function of time using the GC-TDSE.

The initial state in Equation~\ref{first_state} yields $P^k(0) = \delta_{k1}$. This can be seen in the right hand panel of Fig.~\ref{evol_ground}. This value stays constant until the wavepacket hits the constriction at around $t= 2500$ a.u. At this moment, non-adiabatic transitions between different transverse states start to occur and lead to complicated interference patterns at later times (see, e.g., the reduced density $\rho(x,t)$ at $t=5010$ a.u.).
The number of significantly populated transverse states increases up to six (while up to eleven states are required to reproduce the exact reduced density up to a 0.1\% error). Among these states, only odd transverse states are accessible due to the symmetry of the initial state (as we noted in Equation~\ref{even_only}). 
Since the mean energy of the initial state in Equation~\ref{first_state} ($\langle \mathcal{\hat E} \rangle = 0.0037$ a.u.) is higher than the barrier height of the first effective potential-energy in Figure~\ref{ES} (d) ($\max(\mathcal{E}^1_x) = 0.0028$ a.u.), one could naively expect a complete transmission of the wavepacket $\chi^1(x,t)$. However, due to the effect of the non-adiabatic coupling terms $F^{kl}(x)$ and $S^{kl}(x)$, $\chi^1(x,t)$ looses a major part of its population in favour of higher energy transverse components that are reflected by much higher effective potential-energy barriers.

Starting with the second initial state in Equation~\ref{second_state} there are up to seven transverse states populated at the initial time, all of which are again odd states due to the symmetry of the initial conditions (see Figure \ref{evol_gauss}(d)). 
Once the wavepacket hits the constriction at $t=3000$ a.u, states 3, 7 and 9 become more dominant than the previously dominant states 1, 3, and 5. Overall, up to 15 states become important to reproduce the exact reduced longitudinal density within a 0.1\% error. 
Given the above two examples, it seems clear that the specific form of the impinging wavefunction does not play a determinant role in the scaling of the number of relevant transverse states even though it do effect the total number of states $N_e$ required to numerically evaluate Equation \ref{nadse}.

\begin{figure}
\centering
\includegraphics[width=.75\textwidth]{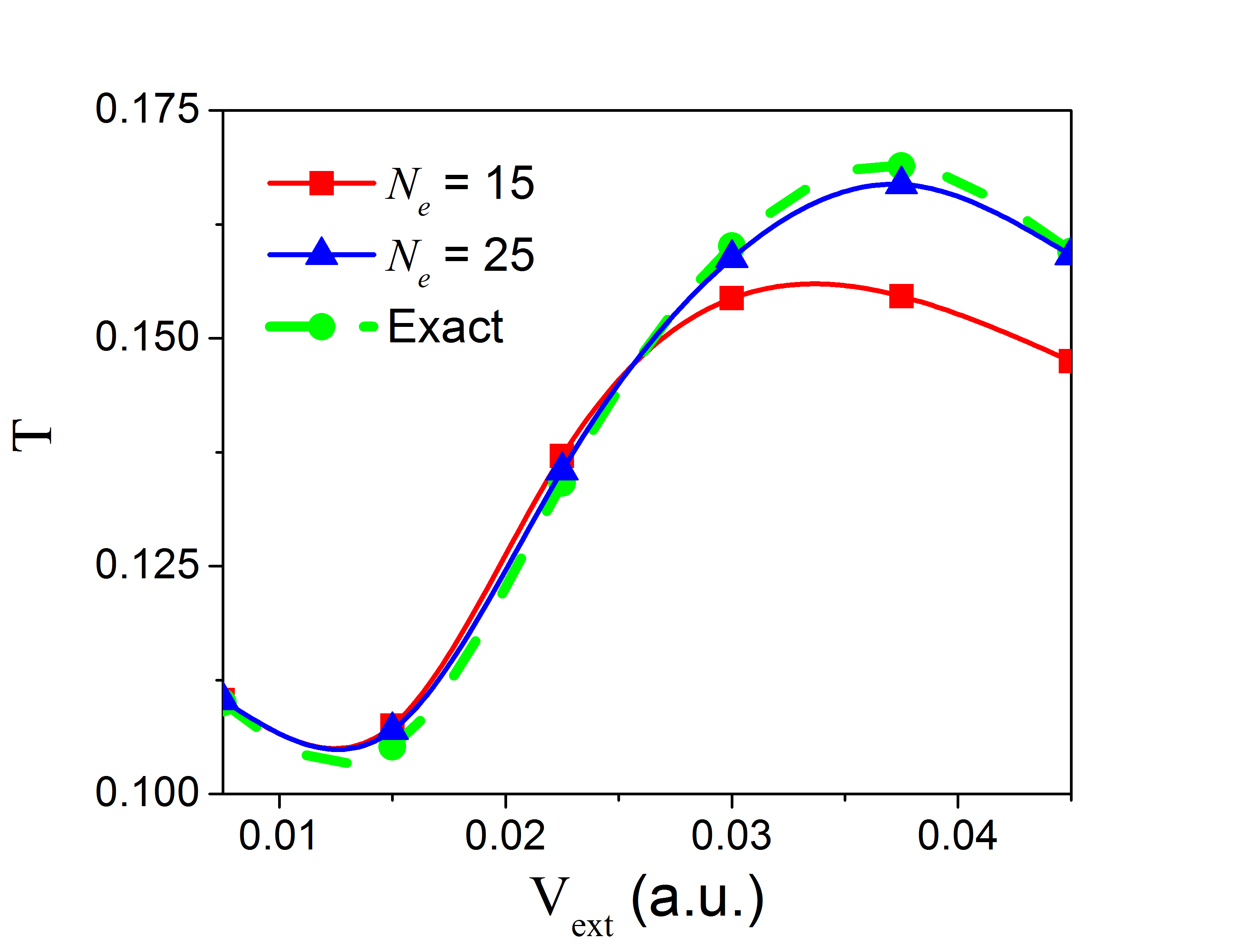}
\caption{Figure depicting the transmission coefficient, $T$, under different bias voltages. This plot gives a comparison of $T$ between the exact 2D simulation (shown in dashed green line) and between the 1D simulation (shown in solid red line for $N_e=15$ states and in solid blue line for $N_e=25$ states). For a voltage range of $0.0075\; \text{a.u.} \leq V_{ext} \leq 0.045\;\text{a.u.}$,  $N_e=25$ states are enough to capture the exact 2D case. For $\text{max}[V_{ext}]=0.027$ a.u.  $N_e=15$ states sufficiently capture the exact 2D case.  }
\label{trans}
\end{figure}

Let us finally consider the effect that an external bias along the longitudinal direction might have on the number $N_e$ of transverse eigenstates required to reproduce the solution of the full 2D TDSE. 
For that, starting with the state in Equation~\ref{second_state}, we consider the transmission coefficient $T = \int_{-\infty}^\infty dy\int_{x_m}^\infty dx  |\Psi(x,y,t_\text{f})|^2$ for different values of the external potential $V_{ext} = V(x)$ in Equation~\ref{appl_ham}. Written in terms of the Born-Huang expansion in Equation~\ref{wf}, the transmission coefficient $T$ reads:
\begin{equation}
    T = \sum_{k=1}^{N_e}\int_{x_m}^\infty dx |\chi^k(x,t_\text{f})|^2,
\end{equation}
where $x_m$ is the center of the nanojunction in the longitudinal direction (i.e., with numerical value $750$ a.u.) and $t_\text{f}$ is the time at which no probability density (i.e., less than 0.1\%) remains inside the constriction.
In Figure~\ref{trans} we show results for applied bias $0.0075\; \text{a.u.} \leq V_{ext} \leq 0.045\;\text{a.u.}$ and two different number of states $N_e = 15$ and $N_e = 25$. As expected, higher applied bias lead to a more vigorous collision of the wavepacket against the constriction due to a higher longitudinal momentum/energy, which allows higher energy transverse states to be populated.       

\section{General discussion}
\label{GD_C}
The GC-TDSE algorithm discussed in the previous sections has a clear computational advantage over the solution of the full 2D TDSE. This is particularly so when the quantities $\mathcal{E}^k(x)$, $F^{kl}(x)$ and $S^{kl}(x)$, involved in the equation of motion of the longitudinal coefficients $\chi^k(x,t)$, are time-independent functions. For a time-independent transverse Hamiltonian, the quantities $\mathcal{E}^k(x)$, $F^{kl}(x)$ and $S^{kl}(x)$ are computed only once before the propagation of the longitudinal coefficients and thus the computational cost of the GC-TDSE resides, mainly, on the propagation of the 1D longitudinal coefficients.  

Let us provide some numbers to get an estimate of the numerical efficiency of the GC-TDSE algorithm. Consider the numerical solution of the full 2D TDSE in a grid. For a number of grid points $\{n_x,n_y\}=\{1500,400\}$,
the resulting Hamiltonian has a dimension $(n_x\times n_y)^2$. Alternatively, the size of the Hamiltonian involved in the propagation of the longitudinal coefficients of the GC-TDSE algorithm is  $(n_x \times N_{e})^2$, where $N_{e}$ is the number of transverse eigenstates. One can then estimate the numerical efficiency of one method over the other by simply evaluating the ratio ${(n_x\times n_y)^2}/{(n_x \times N_{e})^2}$. Thus, for time-independent transverse potentials $W(x,y,z)$, the computational reduction associated to the GC-TDSE is $n_y^2/N_{e}^2$. Note that the benefits of the GC-TDSE would be even more noticeable when applied to a 3D problem, for which the above ratio would become $(n_y^2\times n_z^2)/N_{e}^2$.

As we have seen in the above section, the number of required transverse states $N_e$ is a function of the abruptness/smoothness of the constriction, but also of the energy of the impinging wavepacket. Therefore the computational advantage of the GC-TDSE method over the full dimensional TDSE is clearly system-dependent. Slow wavepackets impinging upon smooth constrictions would maximize the benefits of the GC-TDSE. Contrarily, very energetic electrons colliding against abrupt constrictions would certainly minimize its benefits. In this 
respect, we must note that the GNACs have a clear dependence upon the profile of the constriction. In particular, the second order coupling terms $S^{kl}(x)$ will be sharply peaked for very abrupt constrictions (see the important differences in the size and sharpness of the $S^{kl}(x)$ in Figure \ref{ES} for two different constrictions). Therefore, due to the non-unitary character of the equations of motion of the longitudinal coefficients, very abrupt constrictions may demand very fine grids in practice. 

Finally, let us mention that whenever the transverse Hamiltonian in Equation~\ref{transverse_hamiltonian} is time-dependent, the advantage of the GC-TDSE method compared to the solution of the full dimensional TDSE is not so obvious. As we have already noticed, for a time-dependent transverse potential $W(x,y,z,t)$, the eigenvalue problem in Equation~\ref{incom} must be solved self-consistently with Equation~\ref{nadse}, i.e, at each time step. Then, a comparison of the GC-TDSE and the full dimensional TDSE in terms of numerical efficiency will depend on the specific performance of the eigensolver utilized to evaluate the transverse eigenvalues, $\mathcal{E}^k(x,t)$, and eigenstates $\phi^k_x(y,z)$. 

\section{Conclusions}\label{conclusions}
In this work we have proposed a new method, named GC-TDSE, that allows to include arbitrary geometric correlations between traversal and longitudinal degrees of freedom into a coupled set of 1D TDSE. Our motivation for the development of this method was, initially, a further reduction of the dimensionality of the 3D Schr\"odinger-like equations that result from a (Monte Carlo) SSE approach to quantum electron transport in open systems (valid for Markovian and non-Markovian regimes) that we have recently proposed~\cite{e21121148}. Nevertheless, the method presented here is general and allows to reduce the dimensionality of quantum systems with geometrical correlations among different degrees of freedom, which could be of great utility in different research fields. 

For smooth time-independent constriction profiles under low applied bias, our GC-TDSE method implies up to three orders of magnitude less computational resources than solving the full 3D TDSE directly. 
For very high applied bias or time-dependent constrictions profiles, the GC-TDSE may still be significantly cheaper than the solution of the full 3D TDSE, but would require introducing approximations to the solution of the potential-energies $\mathcal{E}^k(x,t)$ and the GNACs ($F^{kl}(x,t)$ and $S^{kl}(x,t)$). 
We thus expect the GC-TDSE presented here to trigger future investigation for making it robust against more extreme electron transport conditions and to inspire new ways of looking at the many-body problem.












\authorcontributions{Conceptualization, D.P., X.O. and G.A. ; methodology, D.P., X.O. and G.A.; software, D.P.and G.A.; validation, D.P., G.A.and X.O.; formal analysis, D.P., X.O. and G.A.; investigation, D.P., X.O. and G.A.; resources, D.P., X.O. and G.A.; data curation, D.P., G.A., and X.O.; writing--original draft preparation,  D.P., X.O. and G.A.; writing--review and editing, D.P., X.O. and G.A.; visualization, D.P., X.O. and G.A.; supervision, X.O. and G.A.; project administration, X.O. and G.A.; funding acquisition, X.O. }

\funding{We acknowledge financial support from Spain's Ministerio de Ciencia, Innovación y Universidades under Grant No. RTI2018-097876-B-C21 (MCIU/AEI/FEDER, UE). the European Union's Horizon 2020 research and innovation programme under grant agreement No Graphene Core2 785219 and under the Marie Skodowska-Curie grant agreement No 765426 (TeraApps). G.A. also acknowledges financial support from the European Unions Horizon 2020 research and innovation programme under the Marie Skodowska-Curie Grant Agreement No. 752822, the Spanish Ministerio de Economa y Competitividad (Project No. CTQ2016-76423-P), and the Generalitat de Catalunya (Project No. 2017 SGR 348).}


\conflictsofinterest{The authors declare no conflict of interest.} 


\appendixtitles{no} 
\appendix

\section{Derivation of Equation~\ref{nadse} }
\label{non_adiabatic_eq_derivation}
In order to derive Equation~\ref{nadse} we start by introducing the expansion in Equation~\ref{wf} into Equation~\ref{com} to get:
	\begin{equation}
	i \frac{\partial}{\partial t}\sum_k\chi^k(x,t)\phi^k_x(y,z)=T_x\sum_k\chi^k(x)\phi^k_x(y,z)+H(y,z)\sum_k\chi^k(x)\phi^k_x(y,z)+V(x)\sum_k\chi^k(x)\phi^k_x(y,z).
	\label{com2}
	\end{equation}
Making use of Equation~\ref{incom} the above equation can be written as:
	\begin{equation}
	i \frac{\partial}{\partial t}\sum_k\chi^k(x,t)\phi^k_x(y,z)=T_x\sum_k\chi^k(x)\phi^k_x(y,z)+\mathcal{E}^k(x)\sum_k\chi^k(x)\phi^k_x(y,z)+V(x)\sum_k\chi^k(x)\phi^k_x(y,z).
	\label{com3}
	\end{equation}
Expanding the term $T_x\sum_k\chi^k(x)\phi^k_x(y,z)$ as:
	\begin{equation}
		T_x\sum_k\chi^k(x,t)\phi^k_x(y,z)=\sum_k\left[(T_x\chi^k(x,t))\phi^k_x(y,z)-\frac{1}{2}\chi^k(x,t)\frac{\partial^2}{\partial x^2}\phi^k_x(y,z)-\frac{\partial}{\partial x}\chi^k(x,t)\frac{\partial}{\partial x} \phi^k_x(y,z)\right],
		\label{com4}
	\end{equation}
and introducing it back into Equation~\ref{com3} one gets:
	\begin{eqnarray}
	i \frac{\partial}{\partial t}\sum_k\chi^k(x,t)\phi^k_x(y,z)
	&=&\sum_k\left[T_x+\mathcal{E}^k(x)(y,z)+V(x)\right]\chi^k(x,t))\phi^k_x(y,z)\nonumber\\
	&-&\frac{1}{2}\sum_k\left[\chi^k(x,t)\frac{\partial^2}{\partial x^2}\phi^k_x(y,z)+2\frac{\partial}{\partial x}\chi^k(x,t)\frac{\partial}{\partial x} \phi^k_x(y,z)\right]\nonumber\\
	\label{com5}
	\end{eqnarray}
Multiplying both sides of Equation~\ref{com5} by $\int dy \int dz\phi^{*l}_x(y,z)$ and using the orthogonality condition, $\int dy \int dz\phi^{*l}_x(y,z)\phi^{k}_x(y,z)=\delta_{k,l}$ we finally obtain:
	\begin{eqnarray}
	i\frac{\partial}{\partial t}\chi^k(x,t)=\left(T_x+\mathcal{E}^k(x)+V(x)\right)
	\chi^k(x,t) - \sum_{l=1}^\infty\left(S^{kl}(x) +F^{kl}(x)\frac{\partial}{\partial x}\right)\chi^l(x,t),
\end{eqnarray}
where we have defined $S^{kl}=\frac{1}{2}\int dy \int dz \phi^{*l}_x(y,z)\frac{\partial^2}{\partial x^2}\phi^k_x(y,z)$ and $F^{kl}=\int dy \int dz \phi^{*l}_x(y,z)\frac{\partial}{\partial x} \phi^k_x(y,z)$ as the first and second order coupling terms respectively. 

\section{Derivation of Equation~\ref{adiabatic_con} }
\label{adiabatic_condition}
Let us define the wavefunctions $|\phi_l\rangle$ and $|\phi_k\rangle$. Now evaluate the derivative of their inner product as follows,
\begin{equation}
    \langle \phi_l|\phi_k\rangle'=\langle \phi'_l|\phi_k\rangle+\langle \phi_l|\phi'_k\rangle=\delta_{l.k}=0,\;\; k \neq l 
    \label{adiabatic_con2}
\end{equation}
This implies,
\begin{equation}
    \langle \phi'_l|\phi_k\rangle=-\langle \phi_l|\phi'_k\rangle
    \label{adiabatic_con3}
\end{equation}
Evaluating the expression given below,

\begin{equation}
\langle \phi_l|H^\perp|\phi_k\rangle'=\langle \phi'_l |H^\perp|\phi_k\rangle+\langle \phi_l|(H^\perp)'|\phi_k\rangle+\langle \phi_l|H^\perp|\phi'_k\rangle
\label{adiabatic_con4}
\end{equation}
where we have used the chain rule of differentiation. Using the eigenstate-eigenvalue relation from Equation~\ref{incom} we get,

\begin{equation}
\langle \phi_l|H_x^\perp|\phi_k\rangle'=\mathcal{E}^k\langle \phi'_l |\phi_k\rangle+\langle \phi_l|(H^\perp)'|\phi_k\rangle+\mathcal{E}^l\langle \phi_l|\phi'_k\rangle
\label{adiabatic_con5}
\end{equation}
Using the relation in Equation~\ref{adiabatic_con3} in Equation~\ref{adiabatic_con5} we get,
\begin{equation}
\langle \phi_l|H^\perp|\phi_k\rangle'=(\mathcal{E}^l-\mathcal{E}^k)\langle\phi_l|\phi'_k\rangle+\langle \phi_l|(H^\perp)'|\phi_k\rangle=0
\label{adiabatic_con6}
\end{equation}
In the above equation we have made use of the fact that $\langle \phi_l|H^\perp|\phi_k\rangle'=(\mathcal{E}^k\langle \phi_l|\phi_k\rangle)'=0$, when $k \neq l$
Therefore,
\begin{eqnarray}
F^{kl}=\langle\phi_l|(\phi_k\rangle)'&=&\frac{\langle \phi_l|(H^\perp)'|\phi_k\rangle}{\mathcal{E}^k-\mathcal{E}^l}
\label{adiabatic_con7}
\end{eqnarray}
Which can be written in the position representation and using the same nomenclature as used in the main text as follows,
\begin{equation}
F^{kl}(x)=\int dy \int dz \phi^{*l}_x(y,z)\frac{\partial}{\partial x} \phi_x^k(y,z)=\frac{\int dy \int dz \phi^{*l}_x(y,z)\left(\frac{\partial}{\partial x}H^\perp_x(y,z)\right) \phi_x^k(y,z)}{\mathcal{E}^l(x)-\mathcal{E}^k(x)}
\label{adiabatic_con8}
\end{equation}

Using Equation~\ref{transverse_hamiltonian} it is easy to see that only the potential function $W(x,y,z)$ will survive after the partial derivative with respect to the variable $x$. so we can equivalently write Equation~\ref{adiabatic_con8} as,

\begin{equation}
F^{kl}(x)=\frac{\int dy \int dz \phi^{*l}_x(y,z)\left(\frac{\partial}{\partial x}W(x,y,z)\right) \phi_k(y,z)}{\mathcal{E}^l(x)-\mathcal{E}^k(x)}
\label{adiabatic_con9}
\end{equation}

\section{Mean energy in a Born-Huang-like basis}
\label{mean_energy}
The mean energy of a 3D system with a Hamiltonian $H(x,y,z)$ is given by,
\begin{equation}
    \langle  \mathcal{\hat E} \rangle=\iiint dx dy dz \Psi^*(x,y,z)H(x,y,z)\Psi(x,y,z),
    \label{mean_E}
\end{equation}
which in terms of the Hamiltonian in Equation~\ref{H_tot} and the wavefunction expansion in Equation~\ref{wf} can be written as:
\begin{eqnarray}
\langle  \mathcal{\hat E} \rangle&=&\sum_{k,l}^{\infty}\int dx \chi^{*k}(x,t)\int dy\int dz \phi_x^{*k}(y,z)(T_x+V(x))\phi_x^l(y,z)\chi^l(x,t)\nonumber\\
&+&\sum_{k,l}^{\infty}\int dx \chi^{*k}(x,t)\int dy\int dz \phi_x^{*k}(y,z)H(y,z)\phi_x^l(y,z)\chi^l(x,t).
\label{mean_E2}
\end{eqnarray}
Introducingn now Equation~\ref{incom} and Equation~\ref{com3} into Equation~\ref{mean_E2} we finally get:
\begin{eqnarray}
\langle \mathcal{\hat E} \rangle &=&\sum_{k,l}^{\infty}\int dx \chi^{*k}(x,t)\int dy\int dz \phi_x^{*k}(y,z)\bigg(T_x\chi^l(x,t)\phi^l_x(y,z)-\frac{1}{2}\chi^l(x,t)\frac{\partial^2}{\partial x^2}\phi^l_x(y,z) \nonumber\\
&-&\frac{\partial}{\partial x}\chi^l(x,t)\frac{\partial}{\partial x} \phi^l_x(y,z)
+V(x)\chi^l(x,t)\phi^l_x(y,z)\bigg)+\sum_k\int dx \chi^{*k}(x,t)\mathcal{E}^k_x(y,z)\chi^k(x,t)\nonumber\\
&=&\sum_{k}^{\infty}\int dx \chi^{*k}(x,t)T_x\chi^k(x,t)+\sum_k\int dx \chi^{*k}(x,t)\mathcal{E}^k_x(y,z)\chi^k(x,t)+\sum_{k}^{\infty}\int dx \chi^{*k}(x,t)V(x)\chi^k(x,t)\nonumber\\
&+&\sum_{k,l}^{\infty}\bigg[-\frac{1}{2}\chi^{*k}(x,t)\int dy\int dz\phi^{*k}_x(y,z)\frac{\partial^2}{\partial x^2}\phi^l_x(y,z)\nonumber\\
&-&\chi^{*k}(x,t)\int dy\int dz\phi^{*k}_x(y,z)\frac{\partial}{\partial x} \phi^l_x(y,z)\frac{\partial}{\partial x}\bigg]\chi^l(x,t)\nonumber\\
&=&\sum_{k}^{\infty}\int dx \chi^{*k}(x,t)\bigg[(T_x+\mathcal{E}^k(x)(y,z)+V(x))\chi^k(x,t)+\sum_{l}^{\infty}\bigg(-\frac{1}{2}\int dy\int dz\phi^{*k}_x(y,z)\frac{\partial^2}{\partial x^2}\phi^l_x(y,z)\nonumber\\
&-&\int dy\int dz\phi^{*k}_x(y,z)\frac{\partial}{\partial x} \phi^l_x(y,z)\frac{\partial}{\partial x}\bigg)\chi^l(x,t)\bigg]\nonumber\\
&=&\sum_{k=1}^{\infty}\int dx \chi^{*k}(x,t)\left[\left(T_x+\mathcal{E}^k(x)+V(x)\right)
	\chi^k(x,t) - \sum_{l=1}^\infty\left(S^{kl}(x) +F^{kl}(x)\frac{\partial}{\partial x}\right)\chi^l(x,t)\right].
\end{eqnarray}

\section{Function defining the nano-constriction}
\label{Nanoconstriction_function}

For our numerical simulations in Section~\ref{ApplicationI} we considered a prototypical constriction where $L_1(x)$ and $L_2(x)$ in Equation~\ref{potential} are defined as:
\begin{subequations}\label{lx1_lx2}
\begin{equation}\label{lx1}
L_1(x)=\mathcal{A}\Bigg[\bigg(1+\exp\Big(\frac{x-a_1}{\gamma}\Big)\bigg)^{-1}+\bigg(1+\exp\Big(\frac{-x+a_2}{\gamma}\Big)\bigg)^{-1}\Bigg]+\mathcal{B},
\end{equation}
\begin{equation}\label{lx2}
L_2(x)=\mathcal{A}\Bigg[\bigg(1+\exp\Big(\frac{-x+a_1}{\gamma}\Big)\bigg)^{-1}+\bigg(1+\exp\Big(\frac{x-a_2}{\gamma}\Big)\bigg)^{-1}\Bigg]-\mathcal{A},
\end{equation}
\end{subequations}
where $\gamma$ defines the sharpness of the constriction, $\mathcal{A}$ and $\mathcal{B}$ define the maximum and the minimum width of the constriction, respectively $\text{max}[L(x)] = \mathcal{B+A}$ and $\text{min}[L(x)] = \mathcal{B-A}$, and $a_1$ and $a_2$ define the length of the constriction $\mathcal{L}_c=a_2 - a_1$.




\externalbibliography{yes}
\bibliography{bibliography}

\sampleavailability{Samples of the compounds ...... are available from the authors.}


\end{document}